\documentclass[fleqn,10pt]{wlscirep}
\usepackage[utf8]{inputenc}
\usepackage[T1]{fontenc}
\usepackage{color}
\usepackage{svg}
\usepackage{amsthm, amssymb, bm}
\usepackage{mathtools}
\usepackage{wasysym}

    \DeclareMathOperator{\me}{e}
    
    \DeclarePairedDelimiter\ton{(}{)}

    \DeclarePairedDelimiter\mean{\langle}{\rangle}

\title{Mapping job complexity and skills into wages}

\author[1, 2]{Sabrina Aufiero}
\author[1,3,4,5]{Giordano De Marzo}
\author[3*]{Angelica Sbardella}
\author[6,3]{Andrea Zaccaria}
\affil[1]{Dipartimento di Fisica Universit\`a ``Sapienza”, P.le A. Moro, 2, I-00185 Rome, Italy}
\affil[2]{Department of Computer Science University College London, 66-72 Gower St, London WC1E 6EA, UK}
\affil[3]{Centro Ricerche Enrico Fermi, Piazza del Viminale, 1, I-00184 Rome, Italy.}
\affil[4]{Complexity Science Hub Vienna, Josefstaedter Strasse 39, 1080, Vienna, Austria.}
\affil[5]{Sapienza School for Advanced Studies, ``Sapienza'', P.le A. Moro, 2, I-00185 Rome, Italy}
\affil[6]{Istituto dei Sistemi Complessi (ISC) - CNR, UoS Sapienza,P.le A. Moro, 2, I-00185 Rome, Italy.}

\affil[*]{angelica.sbardella@cref.it}

\begin{abstract}
We use algorithmic and network-based tools to build and analyze the bipartite network connecting jobs with the skills they require. We quantify and represent the relatedness between jobs and skills by using statistically validated networks. Using the fitness and complexity algorithm, we compute a skill-based complexity of jobs. This quantity is positively correlated with the average salary, abstraction, and non-routinarity level of jobs. Furthermore, coherent jobs - defined as the ones requiring closely related skills - have, on average, lower wages. We find that salaries may not always reflect the intrinsic value of a job, but rather other wage-setting dynamics that may not be directly related to its skill composition. Our results provide valuable information for policymakers, employers, and individuals to better understand the dynamics of the labor market and make informed decisions about their careers.
\end{abstract}
\begin{document}

\flushbottom
\maketitle
%
%
\thispagestyle{empty}

\section*{Introduction}
This paper proposes an analysis of the structure and relatedness of the US labour market through the construction of a network of jobs and a network of skills, and investigates the relationship between the wages of single occupations and their skill-based coherence and fitness. 
Our empirical strategy relies on the recent stream of literature on Economic Complexity (EC). In particular, we exploit i) the Economic Fitness and Complexity algorithm \cite{tacchella2012new} to compute the fitness of jobs and the complexity of skills; and ii) the approach proposed by Saracco et al. \cite{saracco2017inferring} to validate bipartite network projections and build skills and jobs networks.\\
From its inception, EC was primarily focused on growth forecasting based on the patterns of comparative advantage in the global trade network; in the present day, the EC literature has broadened its scope and object of analysis. It is increasingly focusing on different economic dimensions, especially on local or national systems of innovation \cite{balland2018complex,napolitano2018technology,pugliese2019coherent,straccamore2022will}, with special attention to the green technological potential \cite{barbieri2018specialisation,mealy2017economic,napolitano2022green,sbardella2018green,sbardella2022regional}, and the scientific production of countries or regions \cite{cimini2014scientific,patelli2017scientific,palmucci2020your}. Furthermore, different studies looked at the labour market through Economic Complexity lenses, from ECI \cite{hidalgo2009} and Economic Fitness \cite{tacchella2012new} measures built with sectoral employment and/or wages \cite{caldarola2022structural, fritz2021economic,sbardella2017economic,turco2020knowledge} to linking economic complexity measures with labour market outcomes at the micro-level \cite{adam2021economic}, wage inequality \cite{sbardella2017economic}, labour productivity \cite{basile2019economic}, and the demand for low-skill service jobs \cite{bosio2020economic}. 

The starting empirical tool of the EC studies is usually a bipartite network: here, we focus on the skill-job network obtainable from the Occupational Information Network (O*NET) dataset. In particular, we draw from the measure of relatedness and coherence proposed in the seminal study by Teece et al.  \cite{teece1994understanding} - who studied the relationship between firm performance and the coherence within corporate activity portfolios - to build a measure of job coherence based on each job skill requirements. Moreover, we build on the recent contributions on the product space \cite{hausmann2007structure, hidalgo2007product} and especially from the statistically validated approach called Product Progression Network (\cite{zaccaria2014taxonomy,zaccaria2018integrating,pugliese2019unfolding}; but see also  \cite{tacchella2021relatedness} for a multi-product, non-linear approach based on machine learning). These studies employ export data to measure the relatedness of products through statistically significant patterns of co-exporting in the international trade network and argue that countries that can successfully export a product have developed a set of capabilities or can recombine pre-existing capabilities that enable them to diversify into related goods.
We rely on the Product Progression Network methodology because it has two main advantages with respect to previous efforts \cite{hidalgo2007product}: first, it is dynamical, as it takes explicitly into account time by comparing time-delayed, instead of contemporaneous, co-occurrences; second, it filters each link with a suitable null model, so that only the statistically significant links are considered \cite{cimini2022meta}. The statistical significance of co-occurrences is a critical issue since the presence of a link may not be informative \textit{per se} and could be simply due to the ubiquity of a product or to the diversification of a country/region (here, the degree of jobs and skills). In particular, as in \cite{pugliese2019unfolding,zaccaria2018integrating,cimini2022meta}, in this contribution we make use of the Bipartite Configuration Model \cite{saracco2015randomizing, saracco2017inferring, straka2017grand, vallarano2021fast}, a maximum entropy null-model, to statistically validate the network links.

Evidence of patterns of related diversification is found both at the national and local levels by looking at specialization profiles in industries, products or technologies, highlighting the complementarity among different types of capabilities. A growing body of literature is built on similar approaches to explore different dimensions of relatedness and coherence in terms of knowledge spillovers in a variety of settings, such as industrial sectors (e.g, \cite{neffke2011regions,boschma2013emergence,he2015industry}); activities of plants owned by multi-product firms (e.g, \cite{teece1994understanding}); technological portfolios (e.g \cite{breschi2003knowledge,kogler2013mapping,boschma2014relatedness,napolitano2018technology,rigby2015technological}); also putting into relation different layers of economic activity, such as technological portfolios, exported products and scientific activities \cite{pugliese2019unfolding}, green and non-green knowledge bases \cite{barbieri2022regional} or production \cite{de2022trickle}, through the recently introduced multi-partite networks. 
Our analysis is also closely connected to the notion of skill relatedness put forward by Neffke and Henning \cite{neffke2013skill} who, framing the issues within the Resource-Based View of the firm
\cite{penrose2009theory}, analysed skill linkages across industries by looking at cross-firms labour flows and highlighted the key role of human capital in predicting firm diversification strategies. 
Hartmann et al. \cite{hartmann2019mapping} built a network of jobs using co-occurrences in Brazilian industries, while \cite{muneepeerakul2013urban} defined an analogous network by counting the co-occurrences in metropolitan areas. Landman et al. \cite{landman2022role} used skill-relatedness to study inter-industry spillovers in Irish regions before and after the Great Recession, while in \cite{oclery2022modular} the authors focused on the UK and higher order interactions in labour networks, identifying industry clusters with overlapping skills. Alabdulkareem et al. \cite{alabdulkareem2018unpacking} defined a network of skills using co-occurrences in occupations, connecting the resulting topology to job polarization, i.e. low-skill workers being stuck into a low-wage region of the network. 
Moreover, our study is related to Farinha et al. \cite{farinha2019drives}, who defined three different dimensions of relatedness of jobs using tasks, skills, and co-locations, and studied their impact on exit and entry dynamics in the anatomy of US cities’ employment and occupations. In particular, their notion of skill similarity is close to our measure of skill coherence; however, our focus is on the intrinsic characteristics of jobs and not on local labour markets and the definition and statistical validation of the measures differ. While Farinha et al. defined a measure of job similarity based on \cite{hasan2015lives} using O*NET work activities, we do not have a geographical focus nor do we look at employment level, we instead aim at exploring the intrinsic characteristics of jobs regardless of employment levels, we employ O*NET skills, devise a different digitation criterion and use an entropy-based null model to assess the statistical significance of the links in the network. Furthermore, we define a measure of job economic fitness based on each job skill set, not on the structure of the local labour market, and we investigate the relationship with wages.\\
Our analysis is motivated by several open issues and possible applications.
The empirical evidence shows that human capital mobility is higher between related industries because firms draw upon similar skills \cite{galetti2021skill, maliranta2008labour,neffke2013skill} and thus that labour flows are, among other factors, significantly driven by the similarity in human capital requirements. By shifting the focus from diversification at the industry level to the occupational and skill perspectives, we propose a complementary perspective to that of the skill-relatedness notion put forward by Neffke and Henning. We do not look at labour flows to predict industrial diversification but build a framework able in principle to predict the movement of workers between jobs on the basis of their skill similarity. 
Moreover, we offer a perspective on the structural determinants of occupational wages defining a skill-based measure of job coherence and fitness, study the relatedness among different occupations, examine whether more complex or coherent skill sets map into higher wages and whether our job fitness measure has explanatory power in disentangling the abstract/manual, routinary/non-routinary job-wage relationship. \\
Our measure of job fitness is based on the idea that a complex job requires many complex skills.
We believe that this assessment provides a more fine-grained information on the nature of occupations than the very broad classifications of abstract/manual skills and routine/non-routine tasks, that, according to the mainstream economic consensus, well capture the sophistication of a job. The influential corpus of literature on labour market polarization and routine biased technological change postulates that technological change tends to substitute for human labour in routine tasks -- namely cognitive or manual tasks that can be codified in a finite number of steps and can be easily automated --  and, accordingly, that the falling cost of ICTs penalized medium-skill labour dedicated to routine tasks \cite{autor2006polarization,autor2008trends,goos2007lousy,michaels2014has}. 
These trends led to an increase both in the demand for unskilled labour dedicated to non-routine services and skilled labour devoted to abstract or creative tasks, which are enhanced by ICTs \cite{van2011wage,autor2013growth,autor2015untangling,nedelkoska2015skill}.
However, as argued by \cite{farinha2019drives}, and very much in line with the pursuit of granularity that distinguishes the economic complexity approach, each occupational job class may be a mix of these different broad categories. 
The skill requirements and task composition of each occupation may change over time or across space \cite{deming2018skill,feng2022dynamic}, for different degrees of technology development or adoption may cause context-dependent specificities in the content of the same occupation. Therefore, if the specific task and skill mix within occupations is not fixed, the reliability of labour market polarisation measures and implications may be called into question. 
Contrasting evidence on job polarization was for instance put forth by \cite{autor2013putting}, who found substantial heterogeneity between tasks within the same occupation at the worker level, or by \cite{deming2017growing}, who identified social skills, rather than high skill endowment or sheer job non-routinarity, as a fundamental driver of/the driving force in labour demand. 
Such an aggregated and technologically deterministic view, focusing only on the degree of substitutability between human and automated labour as well as highly aggregated job taxonomies/categorizations/classifications, may also fail to incorporate the tacit knowledge embedded in production lines and not forcefully linked to a codified bundle of skills, or power relationships \cite{cetrulo2020anatomy,dosi2015dynamics} and social and institutional \cite{cetrulo2022vanishing,fernandez2017routine,mishel2017zombie} factors in the organization of work. In this perspective, our assessment of job coherence provides a new tool to quantify how much the skills it requires are similar to each other. To sum up, the Economic Complexity approach offers complementary and alternative instruments to capture these different dimensions and have a more nuanced and granular view of the complexity of each job, how the set of skills that composes it is related, and how it maps into wages. 

\section*{Results}
In this section, we present the results of the application of different tools of the Economic Complexity (EC) methodology to the Skill-Job bipartite network obtained from the O*NET dataset (www.onetcenter.org). This database, fully described in the Methods section, enables us to define an adjacency matrix $\textbf{M}$, whose element $M_{js}$ is equal to 1 if the skill $s$ is relevant for the job $j$ and zero otherwise. Sixty-eight different typologies of skills are reported, while the number of jobs depends on the aggregation level. In particular, we employ two occupational categories: the so-called \textit{Broad Occupations}, comprising a total of 431 different jobs, and the \textit{Minor Groups}, which include 95 job categories. The preliminary data processing approach and the construction of the Skill-Job network are detailed in the Methods section below.

\subsection*{The Skill and Job Progression Networks}
The concepts of relatedness and coherence were originally introduced to study a firm-industry network by Teece et al. \cite{teece1994understanding}. Their aim was to i) reconstruct and quantify the possible overlap of capabilities between industries by counting co-occurrences of products across firms; ii) to use this measure of relatedness (or similarity) between industries to assess the production coherence of firms. As discussed in the introduction, in the EC framework similar approaches are used to assess the relatedness of products by measuring their co-occurrences in the export baskets of countries \cite{hidalgo2018principle} through a network of products, referred to as the product space \cite{hidalgo2007product}. Note that in the EC literature the term \textit{similarity} denotes the affinity between two nodes of the same set of the starting bipartite network (e.g., two products), while \textit{relatedness} usually refers to nodes belonging to different sets (e.g., a country and a product) \cite{tacchella2021relatedness}. Here we do not need this distinction and we will employ interchangeably both terms. In order to quantify the relatedness between pairs of skills and jobs and to build the corresponding networks, we adopt the product progression network approach \cite{zaccaria2014taxonomy,pugliese2019coherent,zaccaria2018integrating}, which also allows for a statistical validation of the resulting network links. The underlying idea behind our approach is that two skills are related if a significant number of jobs require both (for instance, our findings indicate that \textit{Mechanical skills} and \textit{Equipment maintenance}, or \textit{Troubleshooting} and \textit{Quality control analysis} co-occur more often than random). Analogously, two jobs are related if they share a significant number of their skill requirements. By following the methodology proposed by Zaccaria et al. \cite{zaccaria2014taxonomy}, to take into account the nested structure of the adjacency matrix \cite{mariani2019nestedness}, we normalise these simple co-occurrences by job diversification $d_j=\sum_s M_{js}$ and skill ubiquity $u_s=\sum_j M_{js}$. In practice, we compute:
\begin{equation}
	B^{Jobs}_{jj'}=\frac{1}{\max(d_j,d_{j'})}\sum_{s}\frac{M_{js}M_{j's}}{u_s}.
	\label{eq:B_jobs}
\end{equation}
as a measure of the relatedness between job $j$ and job $j'$, and:
\begin{equation}
	B^{Skills}_{ss'}=\frac{1}{\max(u_s,u_{s'})}\sum_{j}\frac{M_{js}M_{js'}}{d_j}
\end{equation}
as a measure of the relatedness between skill $s$ and skill $s'$. \\
The resulting networks, defined by the adjacency matrices $\textbf{B}^{Skills}$ and $\textbf{B}^{Jobs}$, are almost fully connected (practically all possible edges are present) and may contain spurious links, as observed in the country-product case \cite{albora2022machine}. As a consequence, to ensure that only meaningful connections between jobs and skills are considered and spurious links are filtered out, a statistical validation procedure, such as that based on node heterogeneity proposed by Saracco et al. \cite{saracco2017inferring}, is required. As in the normalization procedure, the underlying idea is that a link may be present only because a co-occurring job is highly diversified, or a co-occurring skill is highly ubiquitous; this validation procedure is designed precisely to avoid these occurrences. Further details on the validation strategy are provided in the Methods section.

\begin{figure}[htbp]
    \centering
    \includegraphics[width=0.95\textwidth]{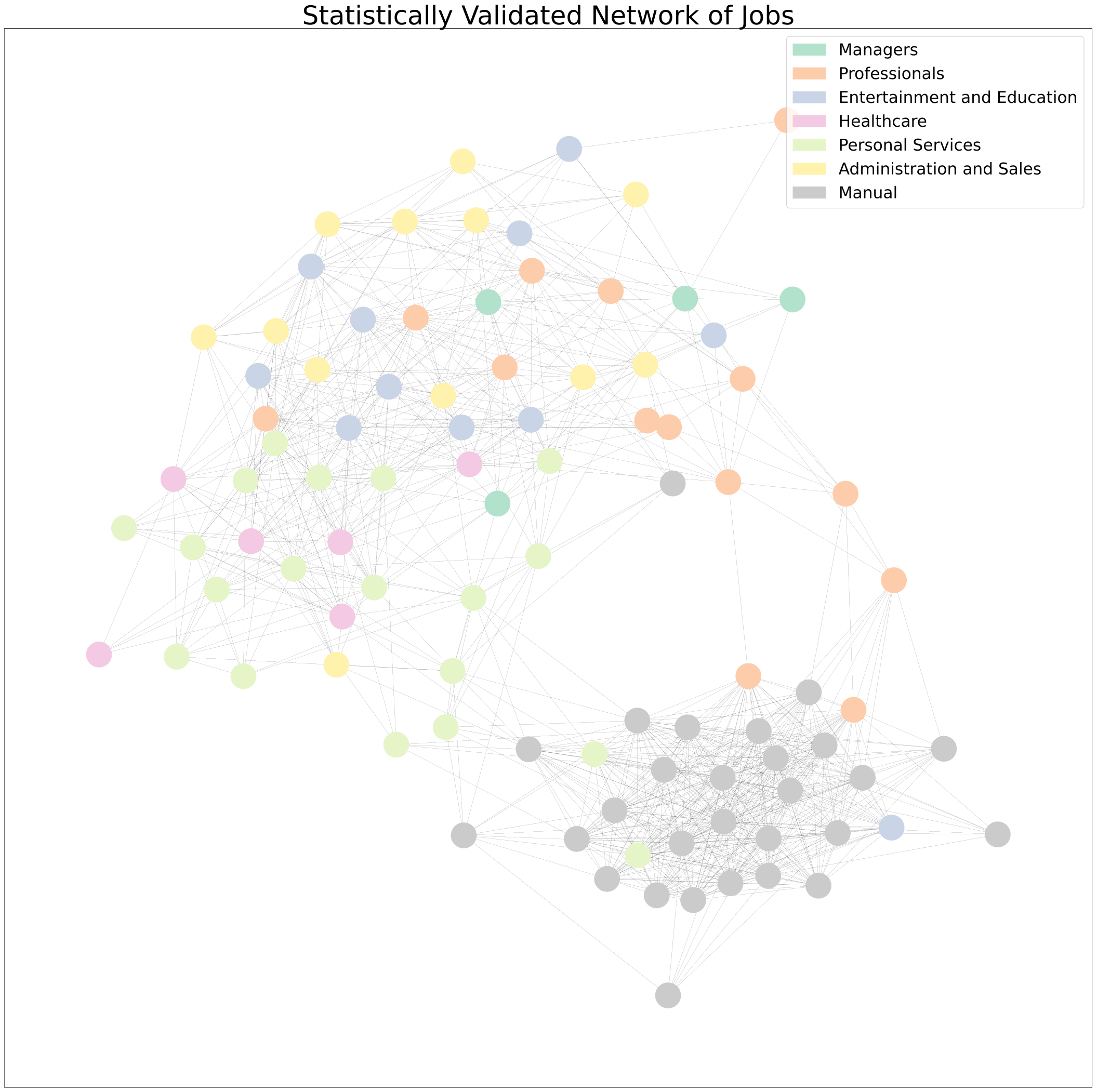}
    \caption{Statistically validated network of jobs. Each node represents an occupation and the links are given by the normalised overlap of skills. A technical/manual job cluster emerges, while abstract occupations appear to be more spread throughout the rest of the network.}
    \label{fig:netjobs}
\end{figure}

The resulting network of jobs is depicted in Fig. \ref{fig:netjobs}, where nodes represent jobs and links are given by normalized and statistically validated co-occurrences of skills.

Intuitively, in this representation jobs that share a relatively high number of skills, i.e. they rely upon similar human capabilities, will be situated close to each other in the network and, once we have filtered the empirical co-occurrences with the null model, proximity indicates a relatively high probability for a worker to change occupation from a job to a neighbouring one. Therefore, its observation allows us to trace the most feasible trajectories for moving to a new occupation on the basis of the shared underlying skills. To illustrate, jobs sharing similar inputs include \textit{Supervisors of food preparation and serving} and \textit{Cooks and food preparation}, or \textit{Agricultural} and \textit{Fishing and hunting}.

We note that O*NET provides a hierarchical classification of jobs that does not correspond with the structure of our network. Colours in Fig. \ref{fig:netjobs} correspond to seven macro-occupational categories obtained aggregating O*NET Major Groups: Managers, Professionals, Entertainment and Education, Healthcare, Personal Service, Administration and Sales, and Manual Occupations. As can be observed in the plot, two clearly separated clusters emerge: the first in the bottom right corner covering manual occupations, and the second in the upper left corner densely connecting different high- and low-skill jobs, ranging from managerial and professional to education, healthcare and personal services, regardless of their position in the classification as can be appreciated by the colour distribution of the nodes. This points out the different and complementary information provided by job relatedness as measured by skill co-occurrences. Indeed, while the O*NET classification reflects more the final occupational requirements and worker attributes, our approach is focused more on \textit{how} the composition of jobs and skills are related to each other: in tune with the economic complexity literature, here we interpret skills are human capabilities, the underlying and connected building blocks necessary to perform a job. \\

More in detail, in the bottom right corner we find a highly connected cluster composed by almost all manual/technical jobs present in the dataset, from construction-related to fishing and hunting occupations. These kinds of jobs share many skills (for instance \textit{Production and Processing}, \textit{Building and Construction}, \textit{Mechanical},  \textit{Equipment Selection}, \textit{Operation Monitoring}), and thus it might be relatively easy to move from one to the other; however, as can be observed in the plot, they are also isolated from other high-skill services and more abstract jobs, highlighting a clear dichotomy in the job landscape and, except for few external nodes, low chances of moving out of the cluster. On the left of the upper cluster, in pink, we find healthcare occupations  -- showing for instance as common skills \textit{Therapy and Counseling}, \textit{Customer and Personal Service}, and \textit{Psychology}. While on the top of the upper cluster are placed administration and sales occupations -- in yellow and characterized by \textit{Administration and Management}, \textit{Education and Training}, and \textit{Economics and Accounting} skills --, managerial and professional occupations (green and orange respectively), sharing the skills \textit{Administration and Management}, \textit{Economics and Accounting}, \textit{Sales and Marketing}, \textit{Critical Thinking}, \textit{Learning Strategies}. Food preparation and serving -- \textit{Personal Services} in lime green -- are more isolated, indicating poor possibilities of retraining due to a high level of specialization in a set of skills hardly useful elsewhere. This network of jobs may constitute a road-map to plan retraining and investment strategies to overcome skill obsolescence and help workers to more easily move to a coherent new occupation, and foster on-the-job training aimed at developing cross-cutting non-task-specific competencies, in particular abstract (cognitive and digital) skills with the aim of diminishing the gap between the two significantly disconnected job clusters.

Next, to better understand the role played by each occupation in the network and the possible connections between the clusters, we quantify the \textit{betweenness} of jobs. This is a measure of centrality that computes the number of shortest paths between all pairs of jobs in the network, displaying high scores for job nodes that act as bridges in connecting different occupational clusters. In particular, in our network we observe two distinct groups of jobs with high betweenness, the first (on the right of the plot) is composed of scientific occupations -- \textit{Computer Occupations}, \textit{Physical Scientists}, \textit{Engineers}) --featuring \textit{Physics}, \textit{Programming}, \textit{Mathematics}, and \textit{Complex problem solving} as common skills. While the second (on the left of the plot) includes manual and technical occupations -- \textit{Motor vehicle operators}, \textit{Cooks and Food Preparation} and \textit{Protective Technicians} -- featuring \textit{Public Safety and Security}, \textit{Law and Government}, \textit{Telecommunications}, and \textit{Monitoring} as common skills.


\begin{figure}[htbp]
    \centering
    \includegraphics[width=0.95\textwidth]{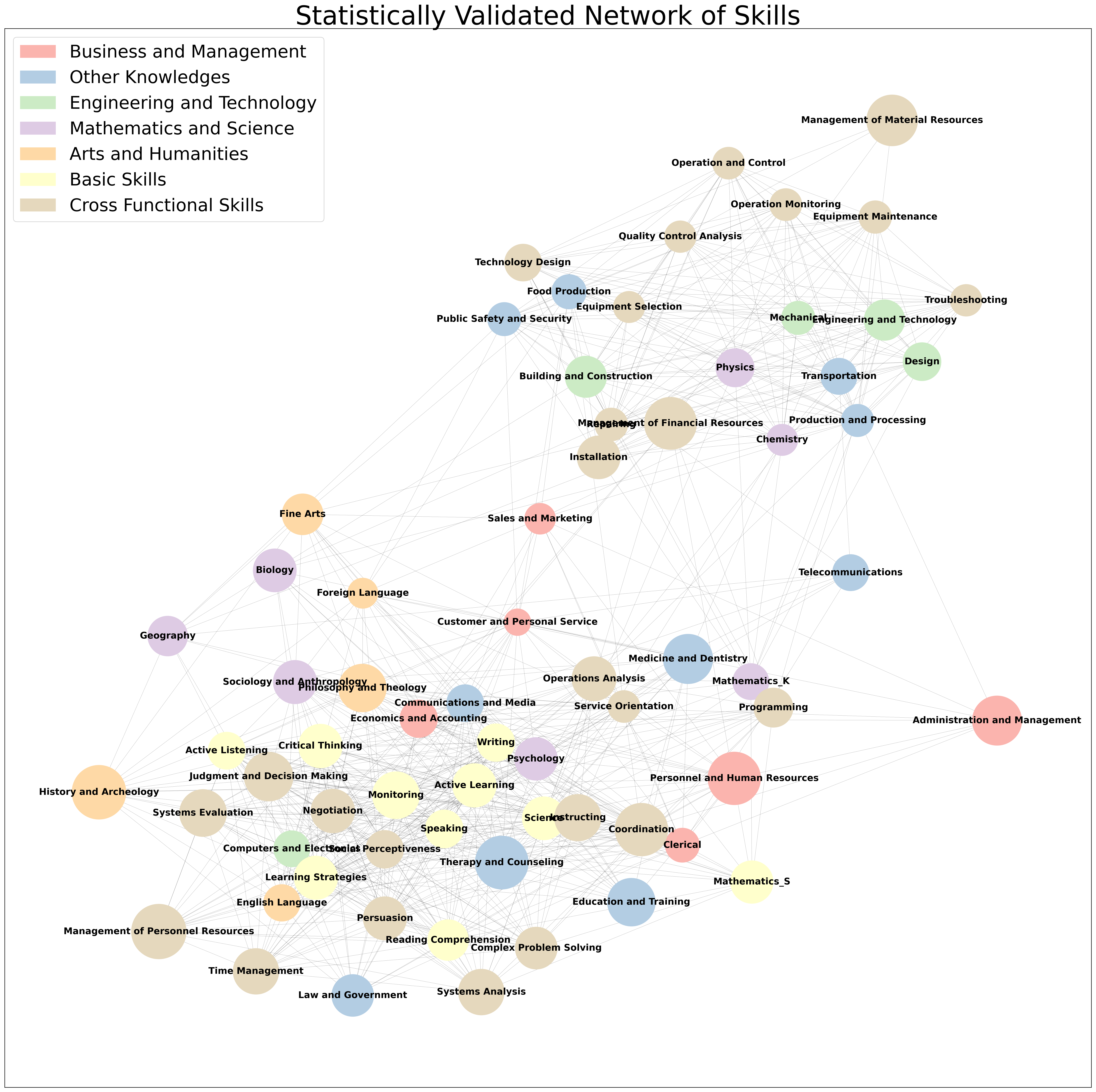}
    \caption{Statistically validated network of skills. Each node represents a skill and the links are given by normalized co-occurrences of skills across jobs. Scientific and technical skills are mainly found in the upper cluster and are mainly independent from humanities and communication skills, found in the lower cluster.}
    \label{fig:netskills}
\end{figure}

In Fig. \ref{fig:netskills} we represent the network of skills, where the colours of the nodes reflect the skill and knowledge classification of O*NET, where two skills are close if a significant number of jobs require both. For instance, examples of highly related skills include: \textit{Mechanical} and \textit{Equipment maintenance}; \textit{Equipment selection} and \textit{Operation monitoring}; or \textit{Troubleshooting} and \textit{Quality control analysis}. 
It is important to notice that in constructing the network of skills we find a strong dependence of the relatedness between skills $\textbf{B}^{Skills}$ from the job classification. At a low aggregation level (Minor Groups, with 95 occupational categories), a large number of unrelated skills co-occur just because the job definition is very wide. This leads to links between skills that are unlikely related, such as \textit{Medicine and dentistry} and \textit{Law and government}, or \textit{Food production} and \textit{Economics and accounting}. To avoid this occurrence and remove such source of noise, when computing $\textbf{B}^{Skills}$, we use the most detailed job classification (\textit{Detailed Occupations}, with 873 occupational categories). This behaviour is analogous to what is observed for product relatedness, which is indeed better tracked at the firm than at the country level \cite{albora2022machine}. 

Also in the case of the skill network in Fig. \ref{fig:netskills} we detect two marked clusters. The first is situated in the top right corner of the network and is associated with industrial production processes being mainly composed of cross-functional technical skills (brown), scientific (purple), and engineering and technical (green) -- almost exclusively represented in this cluster, with the only exception of \textit{Computer and electronics}.  The second cluster can be found in the lower-left corner of the plot. More populated than the first, it densely connects an ensemble of more abstract skills. It covers all basic (yellow) and all art and humanities (orange) skills present in O*NET, while it displays a fair share of cross-functional and scientific skills and health services. Business and management skills (red), e.g. \textit{Sales and marketing}, and \textit{Administration and management}, \textit{Customer care and personal service} together with \textit{Telecommunications} appear to have a tight web of links with both clusters, thus in a way acting as the connecting thread between the two.
 More in detail, the skills composing the first scientific-technical cluster range from fundamental scientific skills such as \textit{Physics} -- which occupies a central position featuring links with almost all the other nodes in the cluster --, \textit{Chemistry}, high-skill technical competences such as \textit{Technology Design}, \textit{Engineering and Technology}, to lower-skill technical competences such as \textit{Installation}, \textit{Construction}, \textit{Mechanics}, and \textit{Equipment Selection}, \textit{Equipment maintenance}, \textit{Operation and control} etcetera. 
 The abstract cluster instead exhibits a combination of basic competencies (e.g., \textit{Speaking}, \textit{Writing}, \textit{Critical thinking} and \textit{Mathematical knowledge}) mainly positioned at the core of the cluster, with different cross-functional skills (e.g., \textit{Social perceptiveness}, \textit{Decision making}, \textit{Problem solving} and \textit{Coordination}). 
 These two appear to be essential inputs for different skills needed in a wide range of professional occupations covering all arts and humanities, some health services such as \textit{Medicine and dentistry} and \textit{Therapy and counseling}, sciences such as \textit{Psychology}, \textit{Biology}, and \textit{Geography}, education, law and communication-related knowledge. As mentioned above, among \textit{Computer and electronics} is the only technical skill disjointed from the first cluster, and interestingly it is densely connected with all basics skills and a large number of the  cross-cutting skills present in the second cluster. This highlights the underlying differences in skills classified in the same category, highlighting complementarity also between apparently unconnected sets of competencies. Our approach thus is  not only able to assess the presence of a skill-based relatedness between jobs but may also provide a quantitative measure to the degree of relatedness between otherwise undetectable links between skills, possibly opening up new developments and applications in designing training or retraining strategies.
 
\subsection*{Jobs complexity, coherence, and wages}
In this section, we discuss an algorithmic assessment of the complexity of jobs based on their skill content. We refer to this metric as \textit{job fitness} and we compute it by applying the Economic Fitness and Complexity (EFC) metric, originally introduced for the country-product bipartite network by Tacchella et al. \cite{tacchella2012new}, to the O*NET data. The underlying idea in constructing this version of the fitness metric is that a job requiring a diversified set of skills, comprising both sophisticated -- i.e., more \textit{complex} -- and basic skills, has higher fitness than a job requiring only a few basic skills. A high-fitness job should require a combination of a wide array of human capabilities and, as we will see in the following, this is only partially reflected in the corresponding wages. 

We compute the fitness of jobs as a function of the complexity of skills and vice-versa by using the EFC algorithm's two coupled iterative equations to define \textit{job fitness} and \textit{skill complexity}. According to the fitness equation, a job is fit if it requires many complex skills. The equation that defines the complexity of skills is, instead, not linear. In fact, high-fitness jobs require both complex and basic skills, thus, in order to determine whether a skill is complex or not, the key information is conveyed by the skill content of low-fitness jobs. If a job presents low human capability requirements, we expect all of the associated skills to be basic. Consequently, if a low-fitness job requires a given skill, that skill will feature low complexity. More details about the mathematical construction and the convergence properties of the algorithm are provided in the Methods sections. In formulas, the EFC algorithm is defined as follows:
\begin{equation}
F_{j}^{(n)}=\sum_{s} M_{j s} Q_{s}^{(n-1)} \\
Q_{s}^{(n)}=\dfrac{1}{\sum_{j} M_{j s} \dfrac{1}{F_{j}^{(n-1)}}}
\end{equation}
where $F_j$ is the fitness of job $j$, $Q_s$ is the complexity of skill $s$, and $n$ is the iteration number of the algorithm. The fixed point of this map does not depend on the initial conditions \cite{cristelli2013measuring} and has non-trivial convergence properties, which are investigated in \cite{pugliese2016convergence}. \\

\begin{figure}[htbp]
    \centering
    \includegraphics[width=0.95\textwidth]{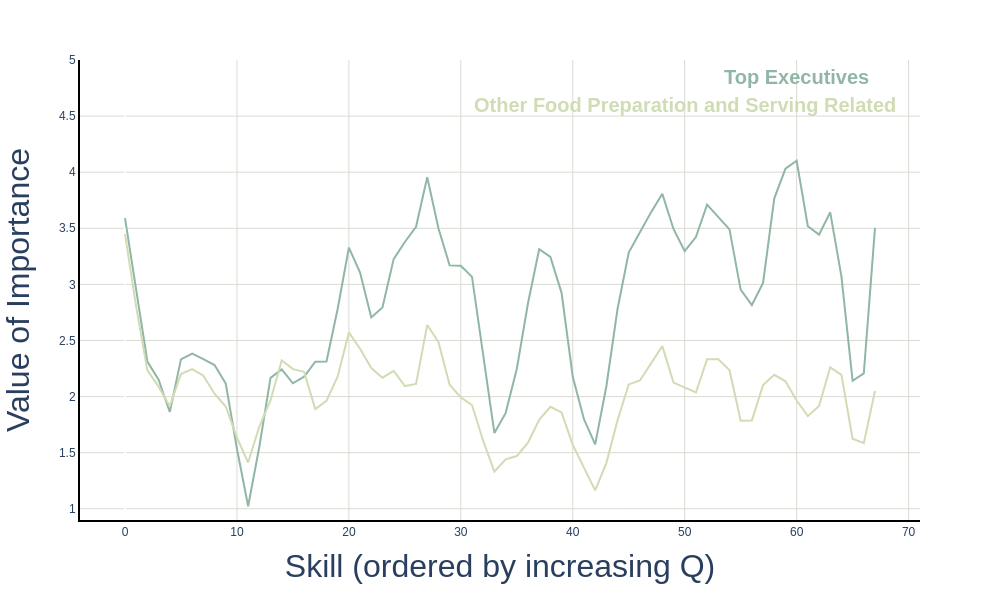}
    \caption{The skill spectroscopy of jobs. The lowest fitness job, represented in light green, requires different skills, but with a low level of importance. By contrast, the highest fitness job, Top Executives in dark green, not only relies on a larger set of complex skills, but these also display high values of importance.}
    \label{fig:spectr}
\end{figure}

The job fitness rankings resulting from this formulation of the EFC algorithm confirm our intuition about the high number of sophisticated skills required in high-fitness occupations. Among the top fitness jobs, we find Management and Top Executives, as well as Supervisors of Repair and Protective Service Occupations, but also Architects and Life Scientists.  By contrast, low-fitness jobs include Food Preparation and Serving Related Workers, Entertainment Attendants, Financial Clerks, Personal Appearance, and Administrative Support workers. By following the graphical representation proposed by \cite{zaccaria2016case} for the \textit{product  spectroscopy of countries}, where each country's exported goods are plotted against their respective complexity, to easily visualise the skill fingerprints of a job it is possible to build a \textit{skill spectroscopy of jobs}. This kind of graphic visualisation is analogous to spectroscopy analysis \cite{bransden2003physics} in materials science, where each material is uniquely identified by the spectrum of the electromagnetic radiation it emits.
For instance, in Fig. \ref{fig:spectr} we provide an example for the skill distribution of one high-fitness job, \textit{Top Executives} in dark green, and that of a low-fitness job, \textit{Food preparation and serving} in light green, i.e., their \textit{skill spectroscopy}. 
On the horizontal axis skills are sorted by increasing complexity -- where for instance  \textit{Quality Control Analysis}, \textit{Customer and Personal Service}, \textit{Equipment Selection} are examples of low complexity skills; while \textit{Therapy and Counseling}, \textit{Coordination}, \textit{Management of Financial Resources} of high complexity skills. On the vertical axis, we report the skill \textit{importance} for each specific job, as detailed in O*NET (see Methods). Note that in O*NET an importance equal to one corresponds to the lowest level of skill mastering. In our example, the importance of each low-complexity skill is very similar for the two selected occupations, as can be observed by the concordant trends followed by the two curves in the left portion of Fig. \ref{fig:spectr}. Moving to intermediate and especially high complexity on the x-axis, we observe a growing discrepancy between the curves, with complex skills appearing as the main driver of the different positions occupied by the two jobs in the fitness ranking.

\begin{figure}[htbp]
    \centering
    \includegraphics[width=0.95\textwidth]{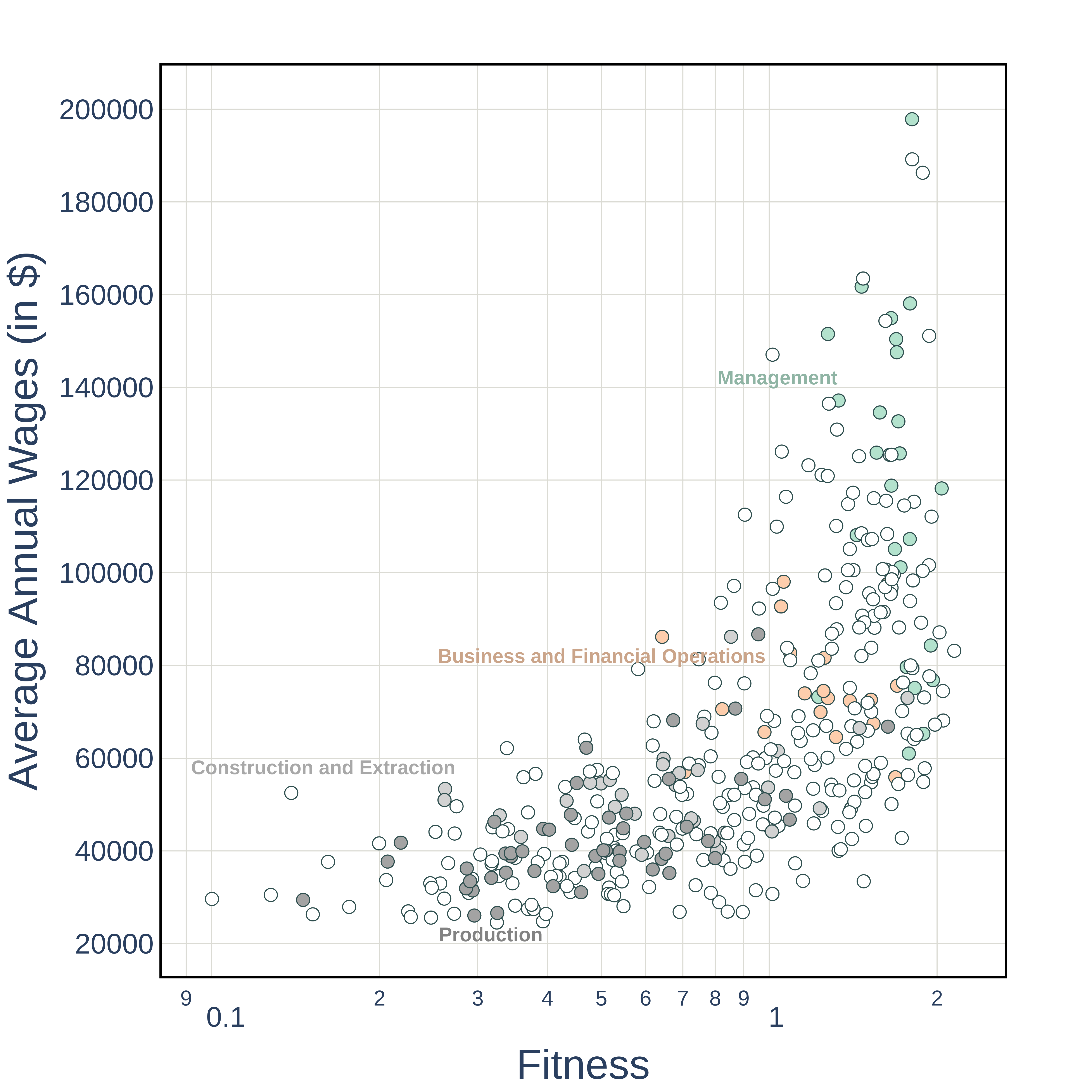}
    \caption{Relationship between job fitness and average wage. Where as jobs we consider O*NET Broad Occupations. The relationship appears to be broadly positive, with a high share of low-fitness jobs confined in the low-wage area. Many high-fitness occupations are however rewarded by heterogeneous wages, independently of their skill content.}
    \label{fig:fitwages}
\end{figure}

Going beyond the sheer dichotomy between abstract vs manual, or routinary vs non-routinary occupations, our fine-grained approach might offer new insights on the skill-wage relationship. Therefore, in Fig. \ref{fig:fitwages} we check how the fitness of jobs, computed as a skill complexity-weighted diversification, is associated to wage levels -- where each scatter plot point represents a job belonging to an O*NET Broad Occupation category.

From the plot it is possible to observe a strong but non linear correlation between fitness and wages, suggesting that the labour market does not always reward occupations with a diversified and complex skill content. For low and intermediate fitness levels, the relationship appears to be broadly positive, with higher skill requirements corresponding to higher average wages. However, noteworthy deviations from such a positive trend are detectable, especially at the two extremes of the wage distribution. First, a share, albeit small, of low-fitness jobs is confined in the low-wage area, with jobs in production or construction (respectively in dark and light grey in the plot) featuring similar low annual average wages despite showing a heterogeneous distribution of fitness -- with e.g. Laundry and Dry Cleaning Workers earning 26600 US dollars per year, and Power Plant Operators, Distributors, and Dispatchers earning 86720 US dollars per year.
Second, for the highest fitness occupations, we detect a much stronger decoupling between a job's complexity skill content and wages, as can be observed by the vertical distribution of the points on the right portion of the plot corresponding to jobs with similar fitness levels and a wide range of high wages, hinting at the emergence of different and relatively skill-independent underlying wage setting mechanisms for very high wages.
This high wage premium is particularly evident in the plot for managerial occupations (in green): managers often share similar skill profiles, and thus fitness, with the bulk of business and financial occupations -- the majority of which is placed in the 60000-80000 US dollars per year region -- however not only they exhibit higher wages but also higher within-occupation wage differentials at similar job fitness levels -- with the extremal cases of Food service Managers earning 61000 US dollars per year, and Engineering Managers earning 158100 US dollars per year. In line with the studies of top income earners \cite{atkinson2011top}, especially for top executives or in financial occupations \cite{mishel2015top}, the ratio of the highest paid manager's salary (Chief Executives, 197840 US dollars per year) to the lowest paid (Food Service Managers) is 3.2; the ratio of the highest paid manager's salary to the lowest paid occupation in general (Fast Food and Counter Workers, 24540 US dollars per year) is 8.06. 
Wages do not always reflect occupations' intrinsic value in terms of skills, they are not simply the outcome of well-functioning competitive markets rewarding skills based on marginal differences: at the very bottom of the pay scale, a sort of wage trap emerges, apparently more driven by the occupation than skill requirements, while when focusing on high-skilled occupations higher wage is accumulated at the very top of the pay scale, regardless of the number or complexity of the specific skills.


\begin{figure}[htbp]
    \centering
    \includegraphics[width=0.45\textwidth]{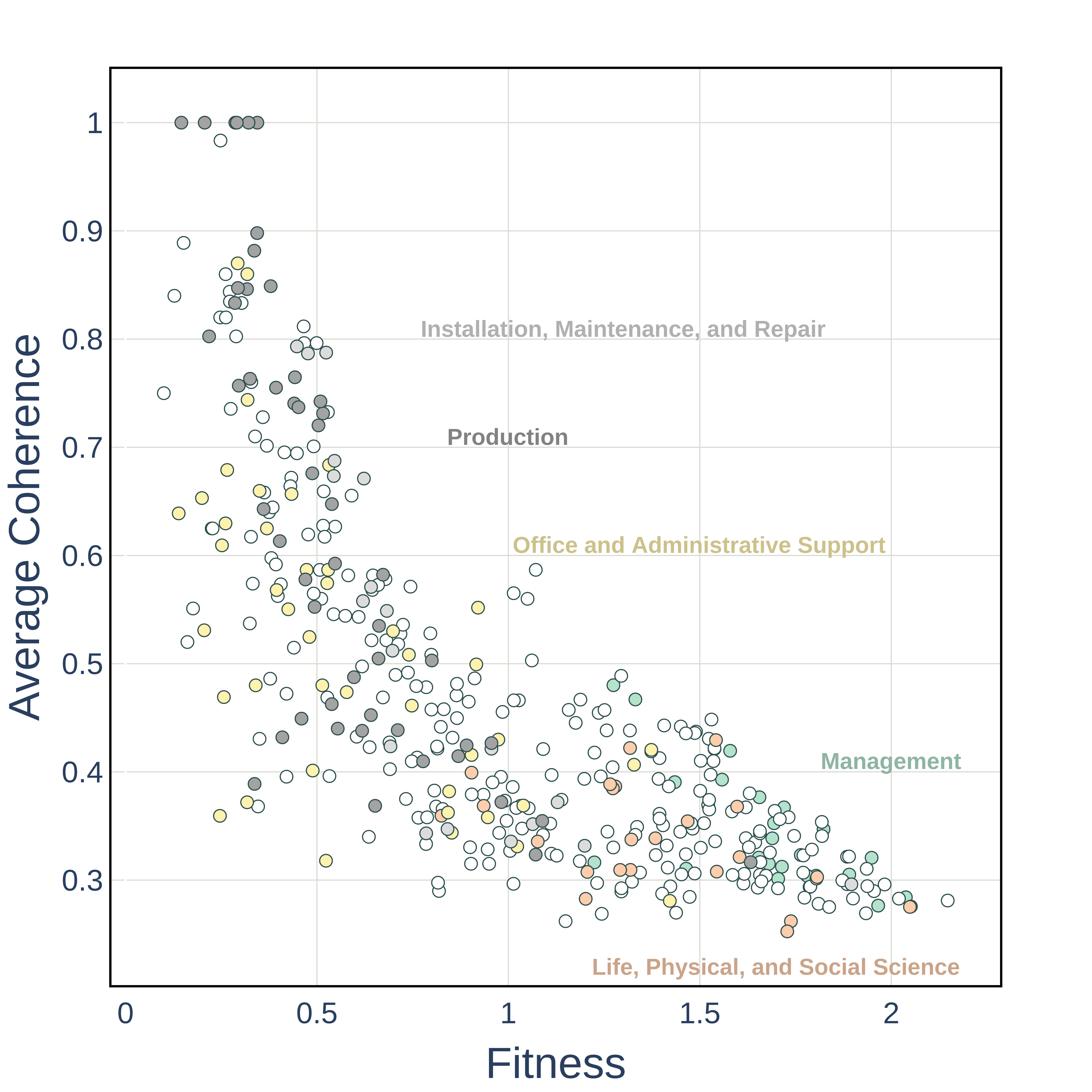}
     \includegraphics[width=0.45\textwidth]{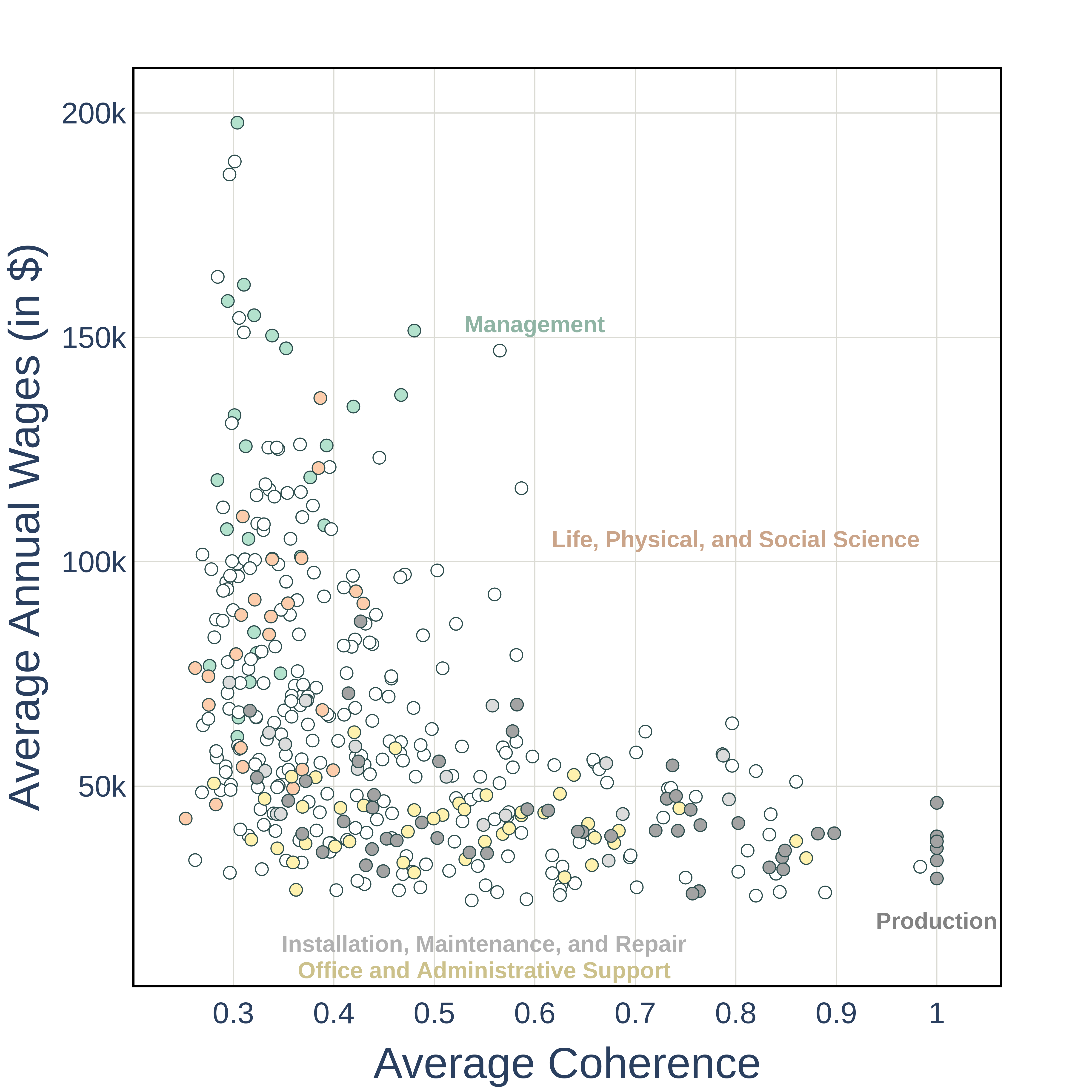}
    \caption{Left panel: Fitness vs Average Coherence (AC) of jobs; right panel: AC vs Wages of jobs. AC and Fitness are mildly negatively correlated. Occupations requiring very similar skills show low wage levels; however, having a low AC is not a sufficient condition for having a high wage level.}
    \label{fig:ac}
\end{figure}
As detailed above and in the Methods section, job fitness measures the degree of sophistication of occupational skill requirements, giving also importance to the skill mix diversification. However, such a measure does not capture the possible role played by the degree of coherence (or lack of thereof) of each job skill requirements, thus their \textit{relatedness}, which can potentially 
 provide additional explanatory power in studying both the relationship with job fitness and wages. To this aim, we exploit the relatedness measure $\textbf{B}^{Skills}$, as introduced in the previous section, to measure the degree of skill coherence of each occupation, that is, the average relatedness between the skills it requires. We denominate this quantity the job Average Coherence (AC) and compute it by averaging the relatedness matrix $\mathbf{B}^{Skills}$ over each pair of skills ${s,s'}$ required by job $j$ according to O*NET, i.e.:
\begin{equation} \label{eq:Gamma}
 \text{AC}_{j} =\frac{\sum_{s s'} {M}_{js} {M}_{js'}B_{ss'}}{\sum_{s s'} {M}_{js} {M}_{js'}} \mbox{ .}
\end{equation}
A job with a high value of AC is \textit{coherent}, i.e. if it requires  more significantly than random co-occurring skills in the same occupations. By contrast, if the required skills are not often associated, the value of AC will be low. Note that this quantity is, by construction, independent from the total number of skills and thus provides complementary information with respect to that provided by fitness.
In the two panels of Fig. \ref{fig:ac}, for each occupation, we show the relationship between the job's AC and respectively fitness (left) and wages (right). From the left panel, it is possible to observe that AC provides a piece of information mainly negatively correlated to fitness: complex jobs have more heterogeneous skill requirements, while low-fitness jobs present a highly coherent skill mix. More in detail, a large share of highly coherent jobs, for instance in the \textit{Installation}, \textit{Maintenance}, and \textit{Repair} and in the \textit{Production} sectors, show low fitness and are characterized by similar skills, consistently with the technical cluster observed in Fig. \ref{fig:netjobs}. Examples include machine operators in \textit{Sewing}, \textit{Woodworking}, and \textit{Molding}. Clerical jobs generally display low fitness values but a relatively high degree of coherence, suggesting that their distinctive skills, albeit relatively simple, are quite dissimilar. 
Finally, \textit{Management} occupations share with \textit{Life, physical, and social scientists} a high value of fitness and a low level of coherence.  Therefore, both categories require many complex skills, and these skills are heterogeneous and unrelated to each other. However, this similar characteristic in terms of skills is not reflected in the distribution of wages, which rewards managerial occupations substantially more. 
In light of the observed relations between fitness and wages, the relationship between AC and average yearly wages in the right panel of Fig. \ref{fig:ac} appears also decreasing but L-shaped with two more extreme emerging patterns. Jobs with low coherence (AC$\lesssim 0.45$) show highly heterogeneous wages, ranging from 30,000 to 200,000 USD dollars with e.g. \textit{Life, physical and social science} occupations in the low-end and managerial occupations in the high-end. By contrast, for jobs with intermediate and high skill coherence (AC$\gtrsim$0.45), e.g. \textit{Different clerical}, and production, installation, and maintenance occupations, it is not possible to obtain wages higher than 60,000 USD dollars, regardless their degree of coherence.

\subsection*{Complex jobs are abstract and non-routinary; coherent jobs are manual and routinary}
The literature on job dynamics and polarization identifies different classes of jobs by using the typology of tasks they usually perform. By following Autor \cite{autor2013growth}, we divide jobs into routinary and non-routinary on the basis of the presence (or lack of thereof) of codified sets of a finite number of simple tasks or procedures, and in manual or abstract if they require manual dexterity and physical strength or intuition and creativity, respectively. 

In this section we explore whether and to what extent the two economic complexity dimensions introduced above, job fitness and average coherence, are connected to these qualitative binary classifications of jobs by highlighting abstract/manual and routinary/non-routinary jobs in the scatter plots presented in Fig. \ref{fig:ac}.

\begin{figure}[htbp]
    \centering
    \includegraphics[width=0.45\textwidth]{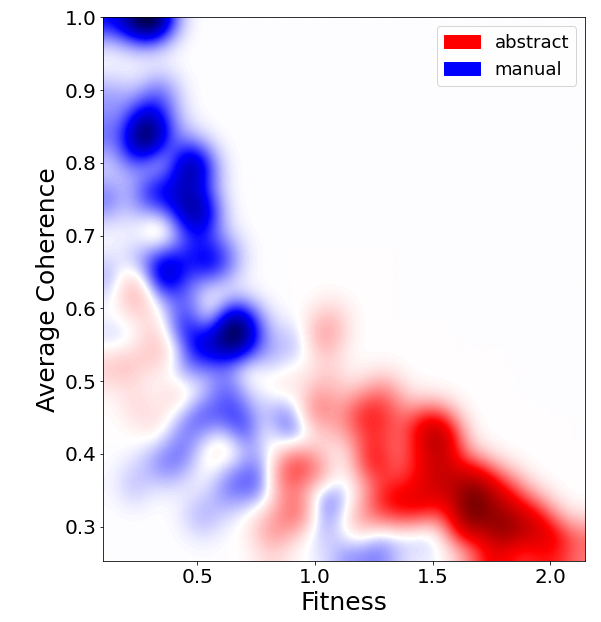}
    \includegraphics[width=0.5\textwidth]{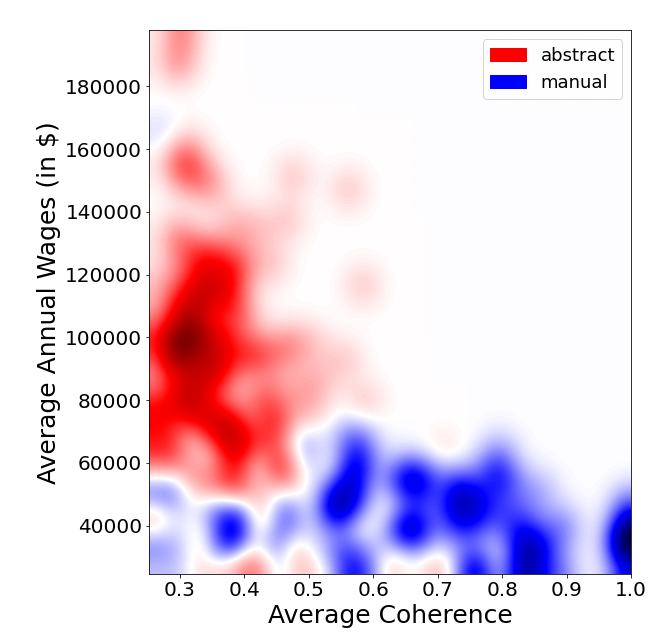}
    \caption{Heatmaps obtained from the scatter plots in Fig. \ref{fig:ac}. Jobs are represented as a function of their average coherence, fitness, and average wages. The colours correspond to their categorization in abstract (red) or manual (blue) occupations. High-fitness jobs are abstract, while highly coherent jobs are manual and present low wages.}
    \label{fig:hm1}
\end{figure}

\begin{figure}[htbp]
    \centering
     \includegraphics[width=0.45\textwidth]{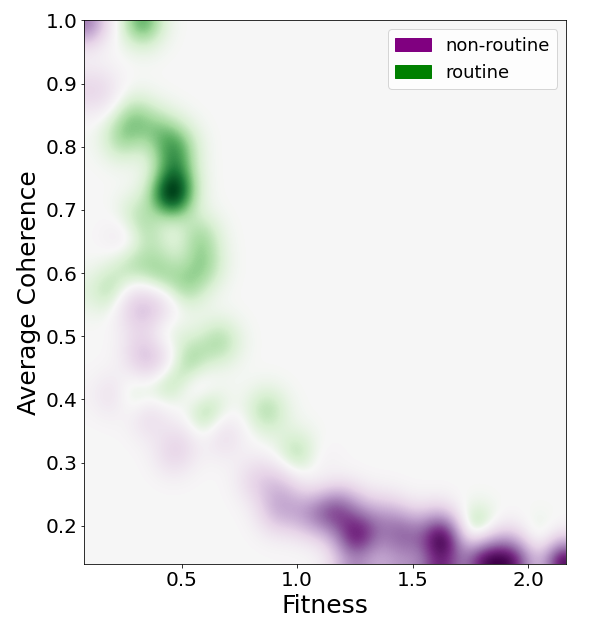}
    \includegraphics[width=0.5\textwidth]{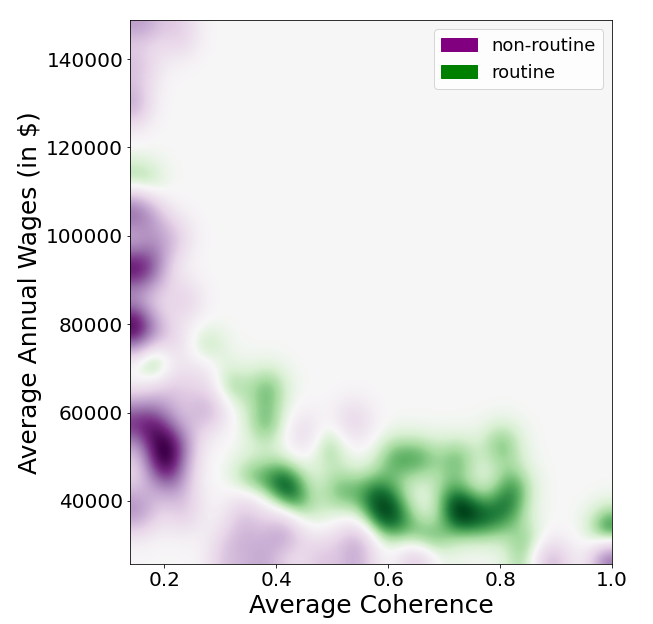}
    \caption{Heatmaps obtained from the scatter plots in Fig. \ref{fig:ac}. Jobs are represented as a function of their average coherence, fitness, and average wages. The colours correspond to their categorization in routinary (green) and non-routinary (purple) occupations. High-fitness jobs are non-routinary, while highly coherent jobs are mostly routinary and present low wages.}
    \label{fig:hm2}
\end{figure}

The heatmaps shown in Fig. \ref{fig:hm1} are obtained by smoothing the two scatter plots shown in Fig. \ref{fig:ac}, in which we coloured the dots in red if the corresponding jobs are abstract, and in blue if manual. The smoothing is performed using a Gaussian kernel (see this \href{https://docs.scipy.org/doc/scipy/reference/generated/scipy.ndimage.gaussian_filter.html}{scipy library}; where we used a value of the kernel parameter \textit{sigma}=32). In such a  way, we are able to visualize the density of job typologies in the plane defined by the Fitness-Average Coherence and Average Coherence-Average Annual Wages. In the AC-Fitness plane in the left panel it is possible to observe that the red abstract area is concentrated in the bottom right corner of the plot, where fitness is high and coherence is low, showing that high-fitness jobs are mainly abstract, while intermediate fitness jobs are a mixture of abstract and manual, and highly coherent jobs are prominently manual. Manual occupations on the top left of the figure include \textit{Electrical and Electronics Assemblers}; \textit{Machine Operators} or \textit{Glaziers}. Examples of abstract jobs (on the bottom right of the figure) are \textit{Secondary School Teachers}; \textit{Fire Inspectors} or \textit{Dentists}. In this area, we find a small number of manual jobs: \textit{Pilots}; \textit{Residential Advisors}; \textit{Boat Operators}. Some abstract jobs, such as \textit{Models},\textit{Billing and Posting Clerks}, and \textit{Telemarketers} have low fitness and low average coherence. From the right panel, we can see that the L-shape of the plot is further characterized when considering the distinction between abstract and manual jobs. Manual jobs occupy the horizontal band of the L-shaped plot: they are characterized by lower wages, below 70,000 USD dollars, and can have any level of coherence, also very low levels, shedding light on the low salaries of the low-coherence jobs in the vertical band. In contrast, how high wages are available only to less coherent, abstract jobs in red in the upper vertical band.\\
In Fig. \ref{fig:hm2} we repeat this exercise, but now the colours correspond to routinary (green) and non-routinary (purple) occupations. Note that given the high class imbalance (the number of non-routinary jobs vastly exceeds the number of routinary jobs) of the Broad Occupations (431 jobs), here we employ the Minor Groups (95 jobs). Looking at the left panel, we observe that the negative relationship between fitness and coherence is here split into two patterns: green in the left upper portion for routine occupations and for non-routine purple in the lower right corner. The high fitness-low coherence area is occupied mostly by non-routinary jobs -- such as \textit{Engineers}; \textit{Postsecondary Teachers}, or \textit{Top Executives}. Routinary jobs are instead characterized by high average coherence and low fitness -- for instance, \textit{Metal Workers}, \textit{Office and Administrative Support Workers}, and Financial Clerks. There are however some deviations to this behaviour: the green points in the bottom right corner, which correspond to technical (supervision) jobs, i.e. \textit{Supervisors of Production Workers}, \textit{Supervisors of Installation},  \textit{Maintenance and Repair Workers}, and \textit{Air Transportation Workers}.
While jobs with high coherence and low fitness are mainly routinary, also here some exceptions can be found, such as \textit{Food and Beverage Serving Workers}; \textit{Grounds Maintenance Workers}; \textit{Entertainment Attendants}, that fall in the non-routinary class. Concerning the right panel, we observe that high wages correspond to non-routinary and low-coherence jobs. With increasing values AC we note a clear lowering of the wages of non-routinary jobs, while routinary jobs with low coherence have relatively higher wages. Finally, low wages and low coherent jobs are mainly non-routinary. 
\\
In conclusion, economic complexity measures can identify different typologies of jobs by extracting relevant information from the skill-job network. This information, such as the complexity content of skills and their coherence, is partially reflected in the abstraction and routinary level of jobs, but, differently from these binary characterizations of jobs, it can be quantified, specifying for instance the degree of complexity for each skill and fitness and coherence for each job. Moreover, AC and fitness both show some interesting deviations from these binary classifications, thus helping us provide additional elements on the relationship between skill, tasks, and wages. Remarkably, non-routinarity is not enough to determine the wage level. Adding a network-based variable accounting for the skill coherence of each job allows us to observe that only high-fitness, low-coherent, abstract, non-routinary jobs are rewarded by high wages. This further stresses the importance of the measures we introduced, since they provide information not contained neither in the wage nor in the routinary/abstraction categorization. 

\section*{Discussion}
The recent social, technological and organisational advancements, the progressive complexification of the world of work and increasing wage inequality in the globalised economy call for new tools beyond standard mainstream economics analysis to investigate and interpret the interplay between human capital, skills and the occupational structure of the labour market. 
The combined availability of increasingly disaggregated data,  algorithmic and complex network techniques allows to unveil novel perspectives and results.
In this paper, by starting from the construction of a bipartite network connecting jobs to the skills they require, we apply different methods from the Economic Complexity toolbox to investigate the structure of occupations and skills in the US labour market relying on the information afforded by the O*NET data-set. 

Firstly, we propose the definition of two statistically validated networks to quantify the similarity between, respectively, pairs of skills and pairs of jobs, useful to understand the possible movement and complementarity of the labour force. The underlying idea behind what we have called the \textit{Network of Jobs} and the \textit{Network of Skills} is that two skills are related if a significant number of jobs require both, and two occupations are related if they share most of their required skills. The statistical validation procedure of network links ensures that only meaningful connections between jobs and skills are considered, and that spurious links are filtered out. The resulting network of jobs allows for tracing the most feasible trajectories for moving to a new occupation on the basis of the shared underlying skills. The network of skills shows two distinct clusters, one including mainly scientific and technical skills and the other soft skills, humanities and communication skills, with different business and management competencies acting as bridges between the two clusters.\\
Secondly, we introduce a novel application of the Economic Fitness and Complexity metric to the job-skill network and propose an algorithmic assessment of the economic fitness of jobs based on the skills they require -- a proxy for their adaptability, resilience, and underlying complexity. The fitness of a job is high if it requires a large portion of complex skills, while low complexity skills are required only by low-fitness jobs. The resulting rankings show that high-fitness jobs include, e.g., management and top executives, architects, and life scientists, but also some specific types of production or administrative occupations, while low-fitness jobs include, e.g., food preparation and serving workers, entertainment attendants, and administrative, maintenance and repair workers. 
Our findings provide evidence of a strong but non-linear correlation between the fitness of a job and its average wage, suggesting that the job market does not always reward occupations with a high number of sophisticated skills. The relatedness between skills can be used to measure the coherence of an occupation: coherent jobs require highly similar skills, i.e. co-occurring more often than random. Our results show that more coherent occupations have lower average wages. In fact, there appears to be a coherence threshold after which it is impossible to go above 60,000 USD yearly wage, while low coherence jobs vary across the occupational classification and display an extremely high variability in wages.
On the one hand, these network-based measures are, a posteriori, closely related to the qualitative binary classification of jobs into abstract and manual, and routinary and non-routinary. High-fitness jobs tend to be abstract and low in coherence, while manual jobs are highly coherent and tend to have lower wages, with the only low-wage low-coherence jobs being manual.
On the other hand, non-routinary jobs have high fitness and span different values of wages.\\
In line with this view and in agreement with a vast empirical literature on the imbalance between the top and bottom wage earners \cite{atkinson2011top,mishel2015top}, our findings challenge competitive models that view salaries as a direct reward for skills. The literature on skills and tasks connects changes in relative demand across occupations and skills mainly to technological change \cite{autor2006polarization,autor2008trends,autor2013growth,goos2009job,michaels2014has}, wherein the substitutability between automated and human labour has been attributed to skills differential firstly and more recently to the job routine-intensity \cite{autor2003skill,acemoglu2017robots,goos2003mcjobs,goos2014explaining}. However, considering (skill/routine biased) technological change as an exogenous unidirectional process may oversimplify the complex interplay between technology, firm dynamics and technology adoption strategies \cite{dosi2000nature,ciarli2021digital}, workers’ bargaining power \cite{dosi2015dynamics,dosi2019whither,cetrulo2020anatomy,van2014tight} and heterogeneity \cite{deming2020earnings,hershbein2018recession}, as well as the institutional factors that contribute to shaping the organization of work \cite{cetrulo2022vanishing,fernandez2017routine,mishel2017zombie}. Wages, therefore, may not simply reflect the intrinsic “value” of work and other wage-setting dynamics not necessarily correlated with job skill and task compositions should also be taken into consideration.

Our findings suggest that economic complexity measures can provide valuable information about different typologies of jobs, allowing for a better understanding of the dynamics of the labour market. In fact, by providing a more nuanced quantitative description of \textit{fitness} and the skill content \textit{coherence} of jobs, our analysis allows to go beyond the dichotomy between manual/abstract and routine/non-routine occupations, and provides additional information on how the structure of the required skill-sets map into wages. Furthermore, by unpacking statistically significant links between pairs of jobs (skills) in the job (skill) network, our analysis allows to trace potential trajectories based on the quantified degree of complementarity or mobility between single jobs (skills) with a high level of detail. 
The findings of this study may inform policymakers and employers on designing more effective labour market policies and on the job-training schemes, that, according to our results, should aim at developing cross-cutting “uncoherent” skill sets that would allow workers to move more easily in the job and skill space aiming at higher wages, and to make more informed decisions about their professional paths.

\section*{Methods}

\subsection*{Database description and preprocessing}
Both the Fitness and Complexity algorithm and the relatedness metrics take binary bipartite networks as inputs. In this section, we illustrate the procedure to obtain the bipartite skill job network from the raw O*NET database.

    \subsubsection*{Connecting skills and jobs}
        We retrieve information on occupations and skills from the Occupational Information Network (O*NET) (www.onetonline.org, www.onetcenter.org), created by the US Department of Labor's Employment and Training Administration (ETA). O*NET provides survey-based information about skills, knowledge, tasks, tools, and technologies connected to each job category, organized according to the O*NET-SOC classification \cite{gregory2019updating}. This classification is hierarchical and contains different levels of aggregation. In particular, the occupations can be organized into:
        \begin{itemize}
            \item Detailed Occupations (873 categories);
            \item Broad Occupations (431 categories);
            \item Minor Groups (95 categories);
            \item Major Groups (22 categories).
        \end{itemize}
        We define our set of skills as the merge of skill and knowledge variables provided by O*NET, obtaining a total of 68 different skills. The skills and knowledge areas are, on average, diffused across all occupational categories, and the resulting bipartite matrix connecting them to occupations is characterized by a nested structure. 
        For each Detailed Occupation and skill, O*NET provides an assessment of the \textit{Importance} of the different skills. This is a discrete variable in the range $[1, 5]$, quantifying the degree of importance of each skill for the job category under consideration (we neglect the other variable \textit{Level} present in the dataset, as it is highly correlated with the importance). We thus obtain a matrix $\mathbf{\tilde{M}}$, whose element $\tilde{M}_{js}$ associates the importance of skill $s$ to job $j$. Starting from this matrix, whose rows correspond to the job categories of the Detailed Occupations, we build the matrices for the less disaggregated O*NET-SOC occupations. In particular, we compute the importance of a given skill $s$ for the aggregated category $k$ as the weighted average importance of skill $s$ in the corresponding Detailed Occupations that are aggregated into category $k$. In the following, we generally refer to these matrices as $\mathbf{\tilde{M}}$ without mentioning the occupation aggregation level.

        \subsubsection*{Job-Skill binary matrix}
        Starting from the matrix $\tilde{M}$ mentioned above, we define a binary matrix $M$ with the same size, but connecting jobs only to their most relevant skills, as all the EC tools we adopt in the analysis require binary matrices as inputs. In order to binarize the matrix we compute for each skill $s$ its average importance in all occupations; then, we set equal to one the entries corresponding to the jobs with importance greater than the average and to zero otherwise. In formulas:
    \[
        M_{js}=
        \begin{cases}
            1\ \text{if} \ \tilde{M}_{js}>\frac{1}{N_j}\sum_j\tilde{M}_{js}\\
            0\ \text{otherwise}
        \end{cases}
    \]
    where we denoted by $N_j$ the total number of occupations. 

        \subsubsection*{Job wages}
        
    The data about occupational wages is obtained from the Quarterly Census of Employment and Wages (QCEW) dataset of the US \textit{Bureau of labour Statistics} (BLS, https://www.bls.gov/cew/), in which occupations are categorized following the SOC classification.

\subsection*{Null model and validation procedure}
    In order to build the monopartite networks of jobs and skills we compute the respective projections of the bipartite matrix $M$. As explained in the Results section, this can be done using Eq.~\eqref{eq:B_jobs}, that is, in the case of the job network: 
    \[
        B^{Jobs}_{jj'}=\frac{1}{\max(d_j,d_{j'})}\sum_{s}\frac{M_{js}M_{j's}}{u_s}
    \]
    where $d_j=\sum_s M_{js}$ is the diversification of job $j$ and $u_s=\sum_j M_{js}$ is the ubiquity of skill $s$. In this way, by computing the normalized co-occurrences of skills across jobs, we obtain the similarity matrix $\mathbf{B}^{Jobs}$ connecting the jobs that share many common skills. Analogously, one can compute a matrix $\mathbf{B}^{Skills}$ whose element $B^{Jobs}_{ss'}$ quantifies the similarity between two skills $s$ and $s'$ in terms of normalized co-occurrences. However, such an adjacency matrix generally corresponds to an almost fully connected network due to spurious co-occurrences, therefore it is necessary to statistically filter the matrix links by exploiting a suitable null model. To this aim, we employ the Bipartite Configuration Model (BiCM) \cite{squartini2011analytical,saracco2015randomizing,saracco2017inferring}, a null model designed to randomize bipartite networks based on the exponential random graph theory. In particular, we rely on the Python BiCM library \url{https://bipartite-configuration-model.readthedocs.io/en/latest/}. The BiCM defines a canonical ensemble of random graphs by constraining (on average) the degree sequences of both node sets (in our case, the ubiquity of skills and the diversification of jobs). We obtain the probability distribution of such an ensemble by maximizing the Shannon entropy under these constraints in the following way:
    \[
        P(\bar{\mathbf{M}}|\{\theta_j\}, \{\mu_s\}) = \frac{\me ^{-H(\bar{\mathbf{M}}|\{\theta_j\}, \{\mu_s\})}}{Z(\{\theta_j\}, \{\mu_s\})},
    \]
    where:
    \begin{itemize}
        \item $\bar{\mathbf{M}}$ is the adjacency matrix of the random bipartite network;
        \item $\{\theta_j\}$ and $\{\mu_s\}$ are the Lagrange multipliers associated respectively to the diversification $\{d_j\}$ and the ubiquity $\{u_s\}$;
        \item $H(\bar{\mathbf{M}}|\{\theta_j\}, \{\mu_s\})$ is the Hamiltonian, defined as:
        \[
        H(\bar{\mathbf{M}}|\{\theta_j\}, \{\mu_s\})= \sum_j \theta_j d_j(\bar{\mathbf{M}}) + \sum_s \mu_s u_s(\bar{\mathbf{M}});
        \]
        \item $Z(\{\theta_j\}, \{\mu_s\})$ is the partition function of the Hamiltonian:
        \[
            Z(\{\theta_j\}, \{\mu_s\})=\sum_{\bar{\mathbf{M}}}\me ^{-H(\bar{\mathbf{M}}|\{\theta_j\}, \{\mu_s\})}.
        \]
    \end{itemize}
    It can be shown that the probability distribution factorizes and takes the analytical form: 
    \[
        P(\bar{\mathbf{M}}|\{\theta_j\}, \{\mu_s\})=\prod_{j, s}p_{js}^{\bar{M}_js}\ton*{1-p_{js}}^{1-\bar{M}_{js}}
    \]
    where $p_{js}$ is the probability that job $j$ and skill $s$ are connected in the random network and satisfies the following condition:
    \[
        p_{js} = \frac{1}{\me^{\theta_i+\mu_j}+1}.
    \]
    While the numerical values of the Lagrange multipliers are obtained by solving the system of equations: 
    \[
        \begin{cases}
            d_j = \sum_s p_{js}\\
            u_s = \sum_j p_{js}.
        \end{cases}
    \]
    
    Once we obtain the probability distribution of the random matrices and, in particular, the probability for the presence of each link in the adjacency matrix of the job-skill network, we can generate $N$ random bipartite networks and, for each of them, compute the projected matrix $\bar{\mathbf{B}}^{Jobs}$ defined above. We then validate the links of $\mathbf{B}^{Jobs}$ by keeping the ones that are larger than the corresponding link in the random matrices in at least the $95\%$ of the cases, while we set to zero the remaining links. Where the statistical significance threshold $95\%$ is arbitrary and sets the confidence level. The same procedure can be applied also to the matrix $\mathbf{B}^{Skills}$ to obtain a validated monopartite network of skills. 
    
\subsection*{Economic Fitness and Complexity algorithm}
    The Economic Fitness and Complexity (EFC) algorithm \cite{tacchella2012new} is an iterative, non-linear map initially introduced to study the Country-Product bipartite network. It allows to compute Fitness, an indicator of the manufacturing capabilities of a country, and Complexity, which quantifies how sophisticated and difficult is to produce a good. The EFC algorithm has been successfully applied in different bipartite networks, focusing on the scientific \cite{cimini2014scientific}, technological \cite{sbardella2018green} or sectoral  production of regions or countries  \cite{sbardella2017economic}, to rank species in mutualistic networks \cite{dominguez2015ranking} or chess players \cite{de2022quantifying}. In the present paper, we apply the EFC algorithm to the Job-Skill bipartite network and compute for each job $j$ a measure of Fitness $F_j$ by counting how many skills it requires and weighting each skill with its complexity. Therefore, in this context, high-fitness jobs are skill diversified and/or require complex skills. For each skill $s$ we compute a complexity measure $Q_s$ by considering the inverse of the number of jobs requiring $s$ and weighting each job with the inverse of its fitness, in such a way, a skill present in a large number of jobs or required only in low fitness jobs will display low complexity. Thus, the non-linear map which defines the $F_j$s and the $Q_s$s is the following:
    \[
        \begin{cases}
            \tilde{F}_j^{(n)} = \sum_s M_{js}Q_s^{(n-1)}\\
            \tilde{Q}_s^{(n)} = \frac{1}{\sum_j M_{js}\frac{1}{F_j^{(n)}}}
        \end{cases}\quad\quad
        \begin{cases}
            F_j^{(n)} = \frac{\tilde{F}_j^{(n)}}{\mean*{\tilde{F}_j^{(n)}}_j}\\
            Q_s^{(n)} = \frac{\tilde{Q}_s^{(n)}}{\mean*{\tilde{Q}_s^{(n)}}_s}
        \end{cases}
    \]
    where we denote with $n$ the iteration index and use as initial condition $Q_s=1$ for all skills. The iteration of these equations leads to a fixed point that has been proven to be stable and non-dependent on initial conditions \cite{cristelli2013measuring}. We stop the iteration of the algorithm when the ranking of job fitness become stable, exploiting the technique described in \cite{pugliese2016convergence}. To this aim, first we define the growth rate $\alpha_c$ of country $c$ as:
    \[
        \alpha_c = \frac{\log\ton*{F_c^{(n)}-\log\ton*{F_c^{(n-2)}}}}{\log\ton*{n}-\log\ton*{n-2}}.
    \]
    Second, we estimate when country $c$ and country $c+1$ will cross exchanging their position in the ordered sequence as:
    \[
        CI_c = n\ton*{\frac{F_c^{(n)}}{F_{c+1}^{(n)}}}^{\frac{1}{\alpha_{c+1}-\alpha_c}}.
    \]
    Notice that we preliminary ordered the countries so that $F_c^{(n)}>F_{c+1}^{(n)}$ and $CI_c$ is well defined only if $\alpha_{c+1}-\alpha_c>0$. Among these well-defined $CI_c$, we define the Minimum Crossing Iteration as: 
    \[
        MCI^{(n)}=\min_c\ton*{CI_c}
    \]
    which indicates the number of iterations needed to observe a variation in the ranking. In the present work, we stopped the iterative algorithm when $MCI$ reached a value of $10^6$. 
\bibliography{sample}

\section*{Acknowledgments}

The authors acknowledge the CREF project "Complessità in Economia".

\section*{Author contributions statement}

G.D.M., A.S., and A.Z. conceived the research.  S.A. performed the numerical investigations. All authors discussed and analysed the results.  All authors wrote, reviewed, and approved the manuscript. 

\section*{Additional information}

The authors declare no competing interests.

\end{document}